\documentclass{ws-procs961x669}            % default, citations in superscript
\begin{document}

CERN-TH-2021-173

\title{Precision Cosmology and Hubble tension in the era of LSS surveys}

\author{G. Fanizza}

\address{Instituto de Astrof\'isica e Ci\^encias do Espa\c{c}o, Faculdade de Ci\^encias\\
da Universidade de Lisboa, Edificio C8, Campo Grande, P-1740-016, Lisbon, Portugal\\
and
\\CERN, Theory Department, CH-1211 Geneva 23, Switzerland\\
E-mail: gfanizza@fc.ul.pt}

\begin{abstract}
We present a fully relativistic framework to evaluate the impact of stochastic inhomogeneities on the prediction of the Hubble-Lema\^itre diagram. In this regard, we relate the fluctuations of the luminosity distance-redshift relation in the Cosmic Concordance model to the intrinsic uncertainty associated to the estimation of cosmological parameters from high-redshift surveys (up to z = 4). Within this framework and according to the specific of forthcoming surveys as Euclid Deep Survey and LSST, we show that the cosmic variance associated with the measurement of the Hubble constant will not exceed 0.1 $\%$. Thanks to our results, we infer that deep surveys will provide an estimation of the the Hubble constant $H_0$ which will be more precise than the one obtained from local sources, at least in regard of the intrinsic uncertainty related to a stochastic distribution of inhomogeneities.
\end{abstract}


\bodymatter

\section{Introduction}
Along this work, we will summarize the results obtained in Fanizza et al. (2021)\cite{Fanizza:2021tuh} and presented at the parallel session {\it ``Status of the $H_0$ and $\sigma_8$ Tensions: Theoretical Models and Model-Independent Constraints''} of the 16th Marcel Grossmann meeting. In particular, we will quantify the cosmic variance concerning the estimation of $H_0$ from forthcoming high redshift survey with limited sky coverage.

Indeed, the best-fit of the CMB data\cite{Aghanim:2018eyx} allows us to
infer the value of few cosmological parameters with the
highest precision achievable so far in cosmology. Among these parameters, the estimation of $H_0$, namely the current expansion rate of the Universe,
has gained great interest for the cosmologists because of an increasing
tension emerging against the value of $H_0$ itself as estimated from local measurements\cite{Riess:2020fzl}.

This tension is almost of $5\sigma$, since CMB measurements provide\cite{Aghanim:2018eyx} a value for $H_0=67.36\pm0.54$ km s$^{-1}$ Mpc$^{-1}$, whereas late time estimations based on local probes, such as Supernovae Ia (SnIa),
return\cite{Riess:2020fzl} $H_0=73.2\pm1.3$ km s$^{-1}$ Mpc$^{-1}$.

Given this tension and working within the conservative framework of $\Lambda$CDM model, our main interest is to understand whether, in view of the forthcoming Large Scale Structure (LSS) surveys, there could be any theoretical bias increasing the standard deviation in order to alleviate the above-mentioned tension. A first attempt to address this point has been provided in Ben-Dayan et al. (2014)\cite{Ben-Dayan:2014swa}, where the effect of velocity dispersion of local SnIa (at redshift lower than 0.1) has been studied and it has been shown that it can introduce a further intrinsic error in the local estimation of $H_0$ of $\sim 1\%$. This result is in agreement with subsequent analysis\cite{Camarena:2018nbr}, leading to an effect which is large but unfortunately not enough to resolve the tension. Our goal here is to extend the analysis to higher redshifts ($z\le 3.85$), including then also lensing effects to the analysis, in order to forecast the estimated precision for forthcoming surveys, just like Euclid Deep Survey (EDS) and LSST. To this aim, we will assume that the new generation of standard candles known as Superluminous Supernovae (SLSNe) will provide a suitable dataset, following what has been recently claimed in Inserra et al. (2021)\cite{Inserra:2020uki}.

In Sect. \ref{sec:local} we present the estimator for the cosmic variance in realistic high redshift surveys. In Sect. \ref{sec:analytic}, we recall and discuss the analytic tool, such as the 2-point correlation function of the luminosity distance-redshift relation and its monopole, needed for our estimation. In Sect. \ref{sec:estimation}, we generate a survey according to the specific of forthcoming surveys and forecast the expected cosmic variance. In Sect. \ref{sec:discussion} our results are summarized and discussed.

\section{Local measurements}
\label{sec:local}
In order to extend the analysis of Ben-Dayan et al. (2014)\cite{Ben-Dayan:2014swa} to higher redshift, we first need to go beyond the linear relation between luminosity distance and redshift $d_L = z/H_0$. To this aim, we then refer to the
 luminosity distance-redshift relation $d_L(z)$ in the homogenous and isotropic $\Lambda$CDM model
\begin{equation}
d_L(z)=  \frac{1+z}{H_0}  \int _{0} ^z \frac{dz'}{\sqrt{\Omega_{m0}(1+z')^3+1-\Omega_{m0}} },
\label{d_L(H)}
\end{equation}
where $H_0$ is indeed the Hubble constant and $\Omega_{m0}$ is the energy density of the matter today. Eq. \eqref{d_L(H)} then provides an estimator for $H_0$ at higher redshifts as
\begin{equation}
H_0 = \frac{1+z}{d_L(z)}  \int _{0} ^z \frac{dz'}{\sqrt{\Omega_{0m}(1+z')^3+1-\Omega_{0m}} } \, .
\label{eq:H0_back}
\end{equation}
By assuming then \textit{a priori} the value of $\Omega_{m0}$ and given that $d_L$ and $z$ can be independently observed, a measurement of $H_0$ can be directly inferred from the high-redshift surveys thanks to Eq. \eqref{eq:H0_back}.

Eq. \eqref{eq:H0_back} provides also the starting point to estimate the intrinsic uncertainty associated to $H_0$ given by the presence of inhomogeneities all around our observed Universe. This is the lowest theoretical uncertainty we can reach, according to the sample of sources we have access to and is usually named {\it cosmic variance}. We hence have that, if we take into account also inhomogeneities in the Universe, the luminosity distance-redshift relation is modified accordingly as
\begin{equation}
\widetilde{d_L}(z,{\bf n})=d_L(z) \left[1 + \delta^{(1)}(z,{\bf n})+ \delta^{(2)}(z,{\bf n})\right] \, ,
\label{eq:dL_pert}
\end{equation}
where ${\bf n}$ is the observed direction of the given source and $\delta^{(1)}$ and $\delta^{(2)}$ are linear and second order corrections to the luminosity distance-redshift relation.

At this point, the inhomogeneous value of the Hubble constant $\widetilde{H_0}$ can be evaluated as
\begin{align}
\widetilde{H_0} \equiv& \frac{1+z}{\widetilde{d_L}(z)}  \int _{0} ^z \frac{dz'}{\sqrt{\Omega_{0m}(1+z')^3+1-\Omega_{0m}} }
\nonumber\\
=&H_0\left[1 - \delta^{(1)} - \delta^{(2)} + \left(\delta^{(1)}\right)^2 \right](z,{\bf n}) \,.
\label{eq:H0_pert}
\end{align}
We notice that $\widetilde{H_0}$ is now a function of $z$ and ${\bf n}$. This is in line with the fact that inhomogeneities may render the estimation of $H_0$ biased by local structures. In principle, these inhomogeneities may affect not only the precision but also the accuracy of $H_0$ and then shift its averaged value within the sample. Here we neglect this effect since the correction to the Hubble diagram due to the non-linearities has been shown to be small in the regime of redshift of our interest\cite{BenDayan:2013gc,Fleury:2016fda,Fanizza:2019pfp}.

Because of these fluctuations, it is possibile to define an estimator for the variance associated to the background value of $H_0$ inferred from a finite survey of $N$ sources observed at positions $(z_i, {\bf n}_i)$, where $i=1,\dots ,N$ runs over the number of the sources.
Indeed, the variance associated to the average value of $\widetilde{H_0}$ inferred from a finite survey of $N$ sources will be then
\begin{align}\label{eq:cov_matrix}
\sigma^2_{H_0}=&\overline{\left( \sum_{i=1}^N\frac{\widetilde{H_{0}}(z_i,{\bf n}_i)}{N}-H_0 \right)\left( \sum_{j=1}^N\frac{\widetilde{H_{0}}(z_j,{\bf n}_j)}{N}-H_0 \right)}
\\
=&\frac{1}{N^2}\sum_{i,j=1}^N\overline{\left( \widetilde{H_{0}}(z_i,{\bf n}_i) \widetilde{H_{0}}(z_j,{\bf n}_j)-H_0^2 \right)}
=\frac{H^2_0}{N^2} \sum_{i,j=1}^N \overline{\delta^{(1)}(z_i,{\bf n}_i)\,\delta^{(1)}(z_j,{\bf n}_j)}\,.\nonumber
\end{align}
where $\overline{\cdots}$ represents the {\it ensemble average} over all the possibile configurations of perturbations. Last equality of Eq. \eqref{eq:cov_matrix} states that our estimator for the cosmic variance of $H_0$ precisely corresponds to the sum of the 2-point correlation function of the luminosity distance-redshift relation over all the possible pairs of sources in the survey. It then links the details of the surveys, such as angular and redshift distributions of the sources, to the theoretical expressions of the inhomogeneities. We also notice that nonlinear terms in Eq. \eqref{eq:cov_matrix} exactly cancel.

In the next section, these general preliminaries will be applied to the case of linear perturbations of the luminosity distance. This will provide the explicit expression for $\sigma^2_{H_0}$ due to all the linear relativistic corrections.

\section{Analytical expressions for leading order effects}
\label{sec:analytic}
Eq. \eqref{eq:cov_matrix} explicitly shows that the cosmic variance $\sigma^2_{H_0}$ is sourced at the leading order only by linear perturbations. It is then enough to take into account the linear relativistic corrections involved in the $\delta^{(1)}$. By restricting our analysis to scalar perturbations as given in the Longitudinal Gauge without anisotropic stress\footnote{This assumption is in line with the fact that we take into account sources located after the decoupling. General expressions in presence of anisotropic stress can be found in Marozzi et al. (2014)\cite{Marozzi:2014kua} for the non-linear luminosity distance-redshift relation and in Fanizza et al. (2018)\cite{Fanizza:2018qux} for the non-linear redshift.}, the relativistic effects occurring at linear level are well-known\cite{Bonvin:2005ps,BenDayan:2012wi,Fanizza:2015swa,Umeh:2014ana}: we have lensing of the photon geodesics along the line-of-sight due to cosmic structures, peculiar velocity (also known as Doppler) due to the free falling motion of the sources within local inhomogeneities and then fluctuations of the gravitational potential around the source position and along the photon geodesics, leading to local and integrated Sachs-Wolfe effects and time delay. However, effects due to gravitational potentials are negligible with respect to lensing and peculiar velocity. We then focus the rest of our discussion only to the lensing and peculiar velocity.

The impact of the these effects on $\sigma^2_{H_0}$ is then given by
\begin{equation}
\frac{\sigma^2_{H_0}}{H_0^2} = \frac{ 1 }{ N^2 }\sum_{i,j} \sum_{E,E'} \int \frac{dk } {k} \mathcal{P} _\psi(k)\mathcal{W}_{Ei,E'j}\,,
\label{DH_(W)}
\end{equation}
where $i,j=1,\dots,N$ and $E$ and $E'$ label the leading relativistic effects such as Lensing ($E,E'=L$) or Peculiar Velocity ($E,E'=PV$). $\mathcal{P} _\psi(k)$ in Eq. \eqref{DH_(W)} is the so-called dimensionless power spectrum of gravitational perturbations in Fourier space
\begin{equation}
\mathcal{P} _\psi(k) \equiv \frac{k^3}{2 \pi^2} |\psi_k|^2 \,.
\end{equation}
We remark that kernels $\mathcal{W}_{Ei,E'j}$ are functions of the (times) redshifts of the $i$-th and $j$-th sources ($\eta_i$) $z_i$ and ($\eta_j$) $z_j$ and their angular separation $\nu\equiv {\bf n}_i\cdot{\bf n}_j$. In this way, the only terms involved in our analysis are the auto-correlations of lensing and peculiar velocity
\begin{align}\label{eq:pure_kernels}
 \mathcal{W}_{PVi,PVj} =& \Xi _ { i } \Xi _ { j } G _ { i } G _ { j } k^2 \left\{\frac{\Delta\eta_i\,\Delta\eta_j(1-\nu^2)}{R^2}j_2 (kR)\right.\nonumber\\
&\left.+\frac{\nu}{3}\left[ j_0\left(kR\right)-2j_{2}\left(kR\right)\right]\right\}(\eta_i,\eta_j,\nu)\\
\mathcal { W } _ { L i,Lj } = & \frac { 1 } { \Delta \eta _ { i } } \frac{1}{\Delta \eta _ { j } }  \int _ { \eta _ { i }} ^ { \eta_o } d \eta \frac { \eta - \eta _ { i }} { \eta_o - \eta } \int _ { \eta _ { j } } ^ { \eta_o } d \eta' \frac { \eta' - \eta _ { j } } { \eta_o - \eta' } \frac{g (\eta )g (\eta')  }{g^2 (\eta_o) } \left[  k^4 H^4 j_4(kR)\right.\nonumber\\
&\left.- 8 k^3 H^2 L\,j_3(kR)
+ k^2 \left( 8 L^2 -6 H^2 \right) j_2(kR) + 4\,k\,L\,j_1(kR) \right]\left( \eta,\eta',\nu \right)\,,\nonumber
\end{align}
and their cross-correlation
\begin{align}
\mathcal{W}_{ PVi, L j } =&\frac{\Xi_i}{\Delta \eta_j} G_i \int_{\eta_{j}}^{\eta_o}  d\eta \frac{ \eta - \eta_{j} }{ \eta_o - \eta }\frac{g(\eta)}{g(\eta_o)}
\left\{ -k^3 \Delta\eta^2 \frac{(\Delta\eta_i - \nu \Delta\eta) (\Delta\eta - \nu \Delta\eta_i)^2 }{ R^3 } j_3(kR)\right. \nonumber\\
&+ k^2 \Delta\eta \left(3 \Delta\eta \frac{\Delta\eta_i - \nu\,\Delta\eta } { R^2 } - 2 \nu \right) j_2(kR)\nonumber\\
&\left. - \frac{k\Delta\eta}{R} \left[ k^2 \Delta\eta (\Delta\eta_i-\nu \Delta\eta)-2 \nu \right] j_1(kR)\right\} \left( \eta_i,\eta,\nu \right)\,,
\label{eq:mixed_kernels}
\end{align}
where $j_n(x)$ are the spherical Bessel functions of $n$-th order and we define
\begin{align}
R(\eta_x,\eta_y,\nu)=&\sqrt{\Delta\eta^2_x+\Delta\eta^2_x-2\Delta\eta_x\Delta\eta_y\nu}
\nonumber\\
L(\eta_x,\eta_y,\nu) = \frac{\Delta\eta_x\,\Delta\eta_y \nu}{R\left( \eta_x,\eta_y,\nu \right)}\qquad,&\qquad
H(\eta_x,\eta_y,\nu) = \frac{\Delta\eta_x\,\Delta\eta_y \sqrt{1-\nu^2}}{R\left( \eta_x,\eta_y,\nu \right)}\nonumber\\
G_i=\int_{\eta_{in}}^{\eta_i}d\eta \, \frac{a(\eta)}{a(\eta_i)} \frac{g(\eta)}{g(\eta_o)}\qquad,&\qquad
\Xi_i = \left(1-\frac{1}{\mathcal{H}_i \Delta \eta_i} \right)\,,
\label{eq:important_eqs}
\end{align}
with $\Delta\eta_x\equiv\eta_o-\eta_x$. In Eqs. \eqref{eq:important_eqs}, $R$ is the Euclidean distance between two sources and $L$ and $H$ are respectively the normalized scalar and (modulo of the) vector products between the two directions of the sources.
Moreover, $G_i$ is the growth factor for the velocity potential of the $i$-th source, $g(\eta)$ is the growth factor of the gravitational potential, $a(\eta)$ is the scale factor, $\mathcal{H}(\eta)=a'(\eta)/a(\eta)$ and $\eta_o$ is the value of the conformal time today. The detailed analytic derivation for all the different contributions $\mathcal{W}_{Ei,E'j}$ is reported in Fanizza et al. (2021)\cite{Fanizza:2021tuh}. We just remark that the kernel
$\mathcal{W}_{PVi,PVj}$ is in agreement with the one found in Ben-Dayan et al. (2014)\cite{Ben-Dayan:2014swa}.

\begin{figure}
\centering
\includegraphics[scale=0.5]{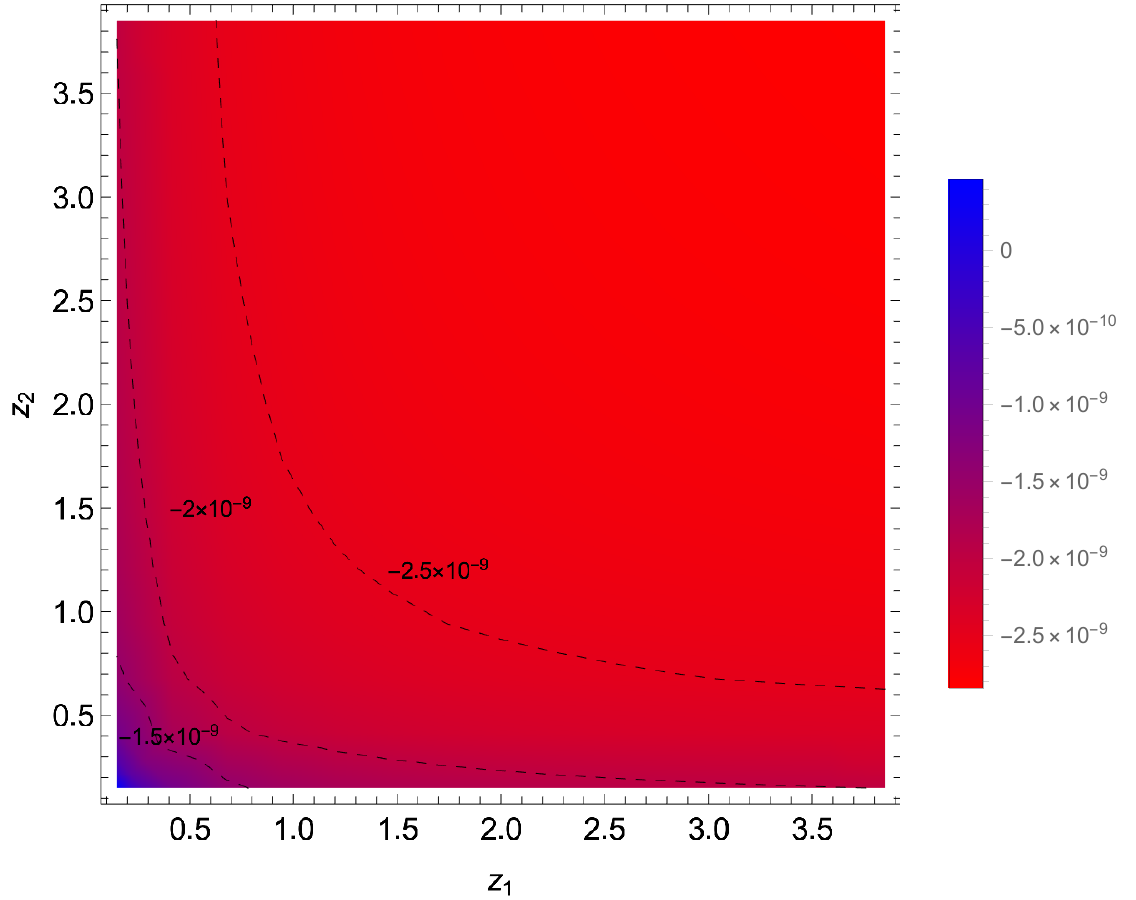}
\includegraphics[scale=0.5]{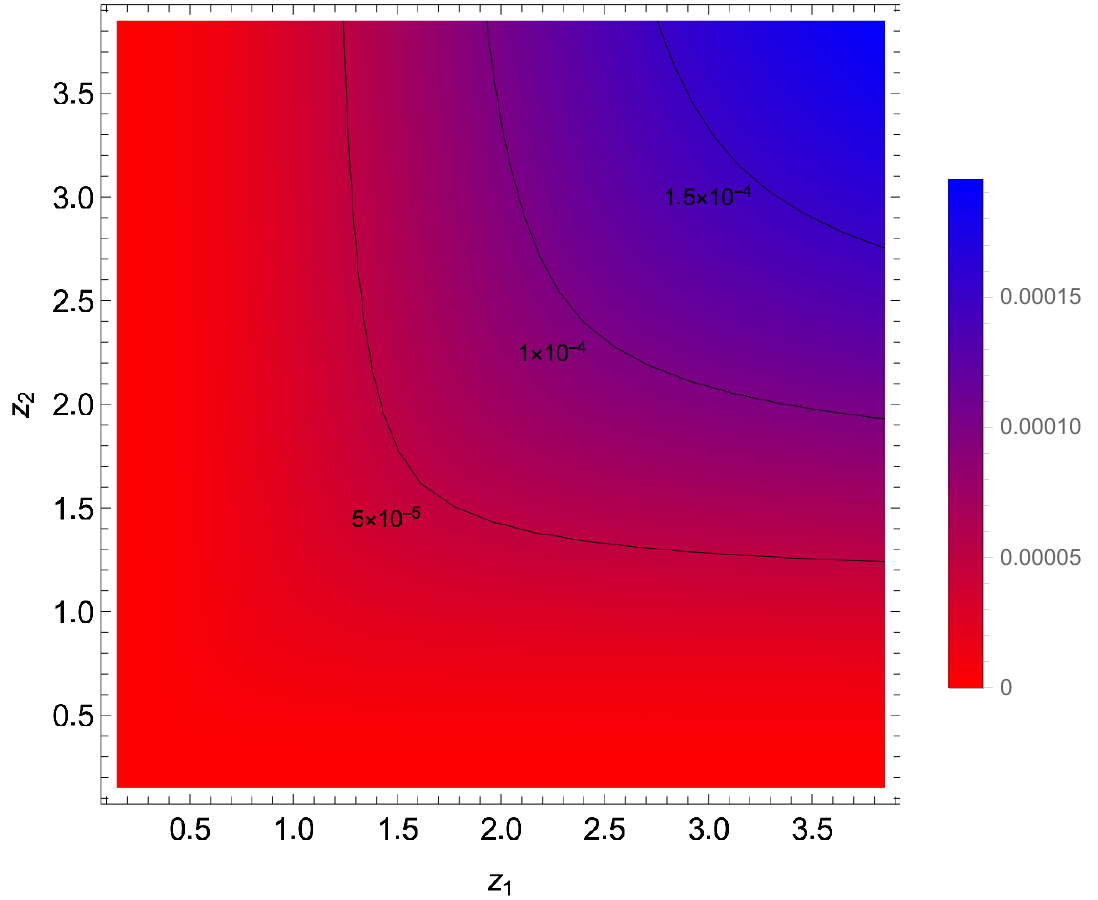}\\
\includegraphics[scale=0.5]{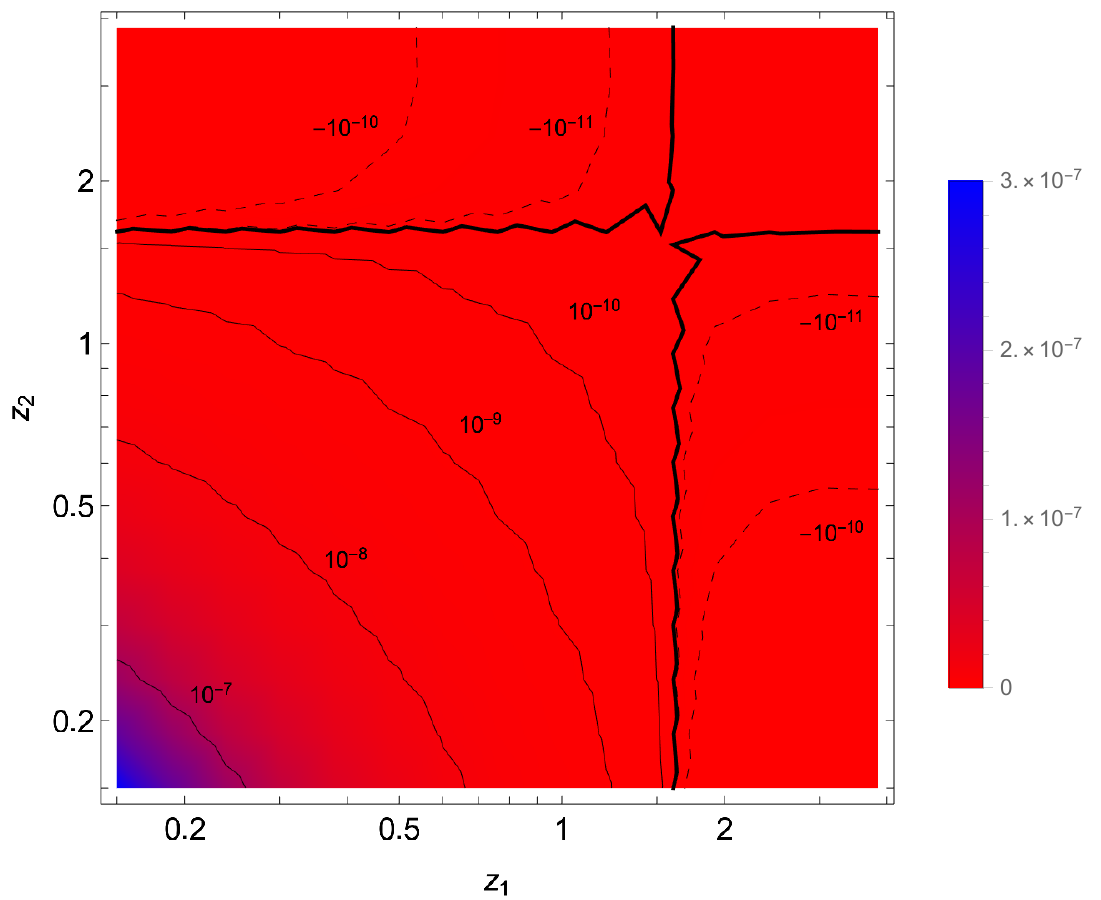}
\includegraphics[scale=0.5]{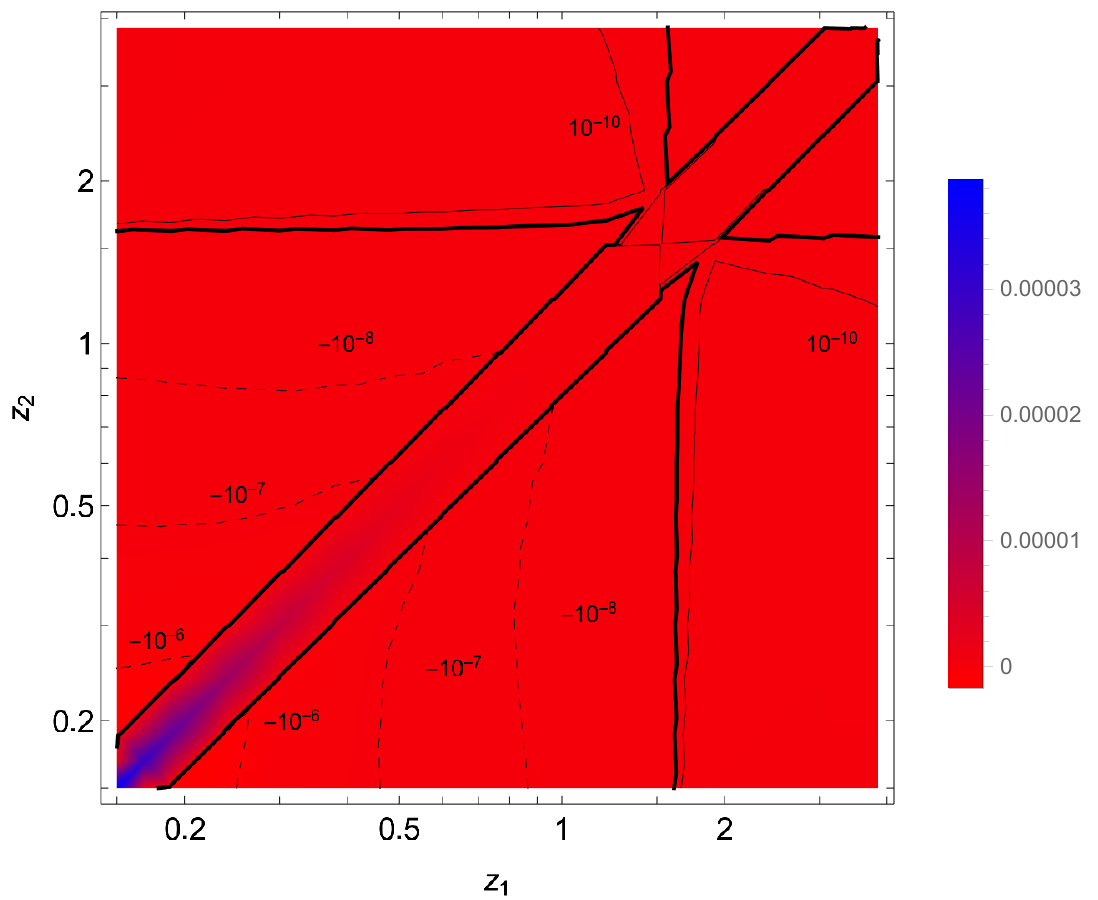}\\
\includegraphics[scale=0.5]{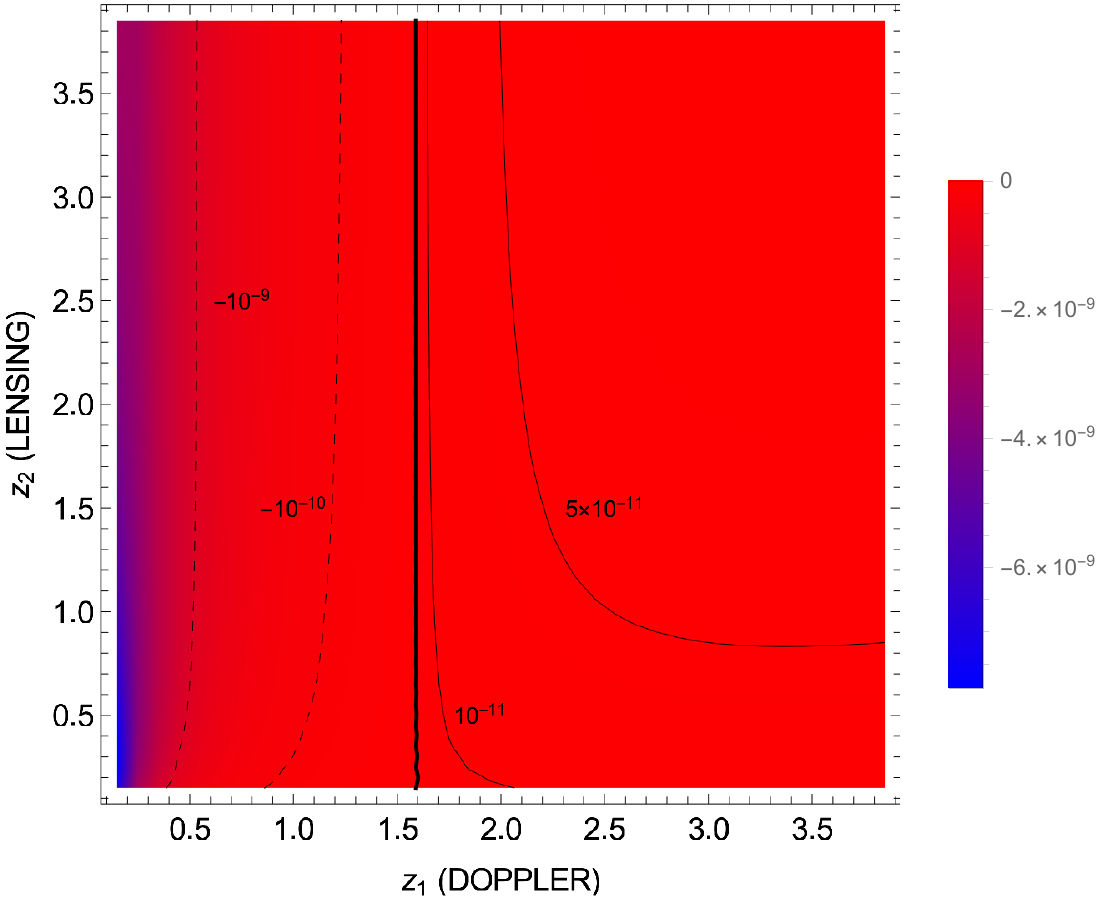}
\includegraphics[scale=0.5]{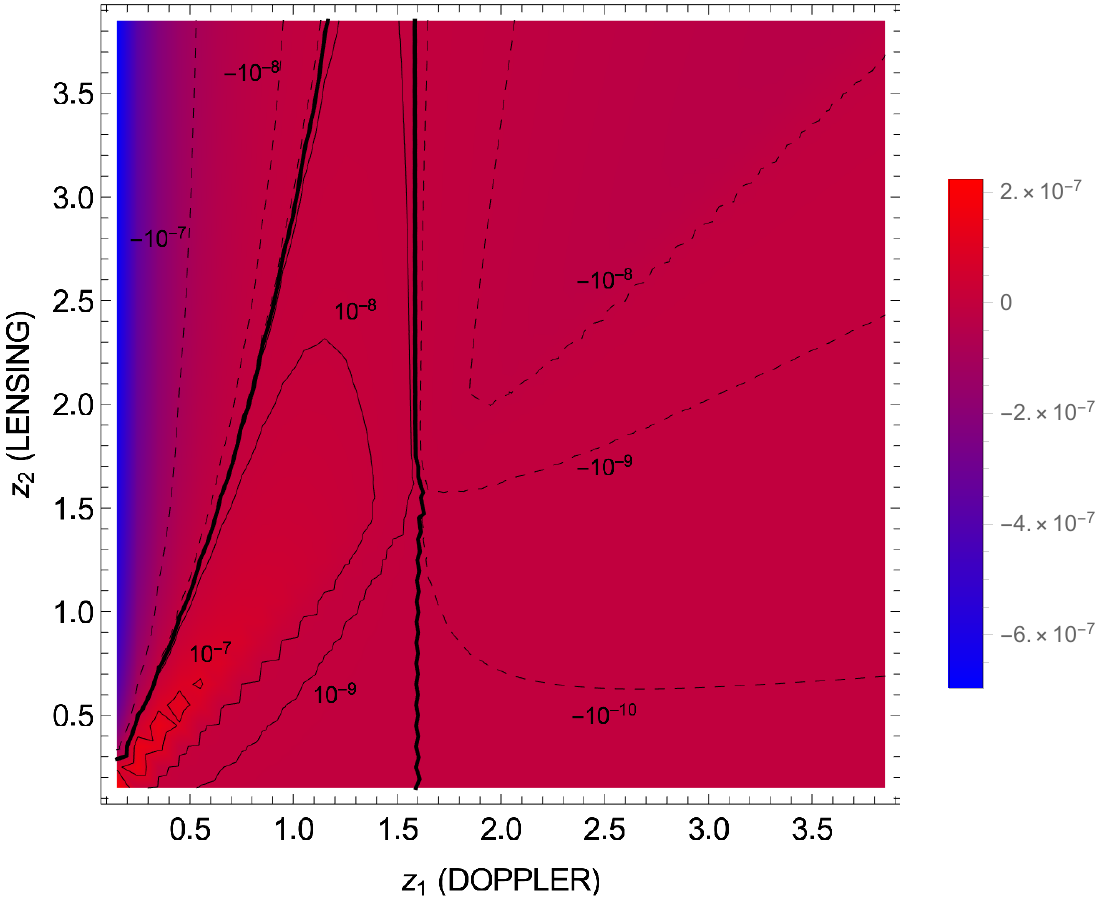}
\caption{2-point correlation functions of Lensing (top panels), Peculiar Velocities (middle panels) and cross-correlation between Lensing and Peculiar Velocity (bottom panels). Left panels refer to anti-aligned sources ($\nu=-1$) wheres right ones consider aligned sources ($\nu=1$). These plots are obtained by using linear power spectrum for the gravitational potential, following the cosmological parameters provided in Fanizza et al. (2021)\cite{Fanizza:2021tuh}.}
\label{fig:amplitudes}
\end{figure}

The values of the 2-point correlation function
\begin{equation}
\xi_{E,E'}(z_i,z_j,\nu)\equiv\int \frac{dk } {k} \mathcal{P} _\psi(k)\mathcal{W}_{Ei,E'j}
\end{equation}
for the kernels in Eqs. \eqref{eq:pure_kernels} and \eqref{eq:mixed_kernels} are then shown in Fig. \ref{fig:amplitudes} for the particular cases of aligned ($\nu=1$) and anti-aligned ($\nu=-1$) sources.
In the regime of interest for us ($0.15\le z\le 3.85$), it is clear from these plots that lensing is always leading with respect to the other terms when the sources are aligned, whereas peculiar velocity turns out to be the leading effect for the anti-aligned sources at redshift smaller than 1.

\begin{figure}[ht!]
\centering
\includegraphics[scale=0.8]{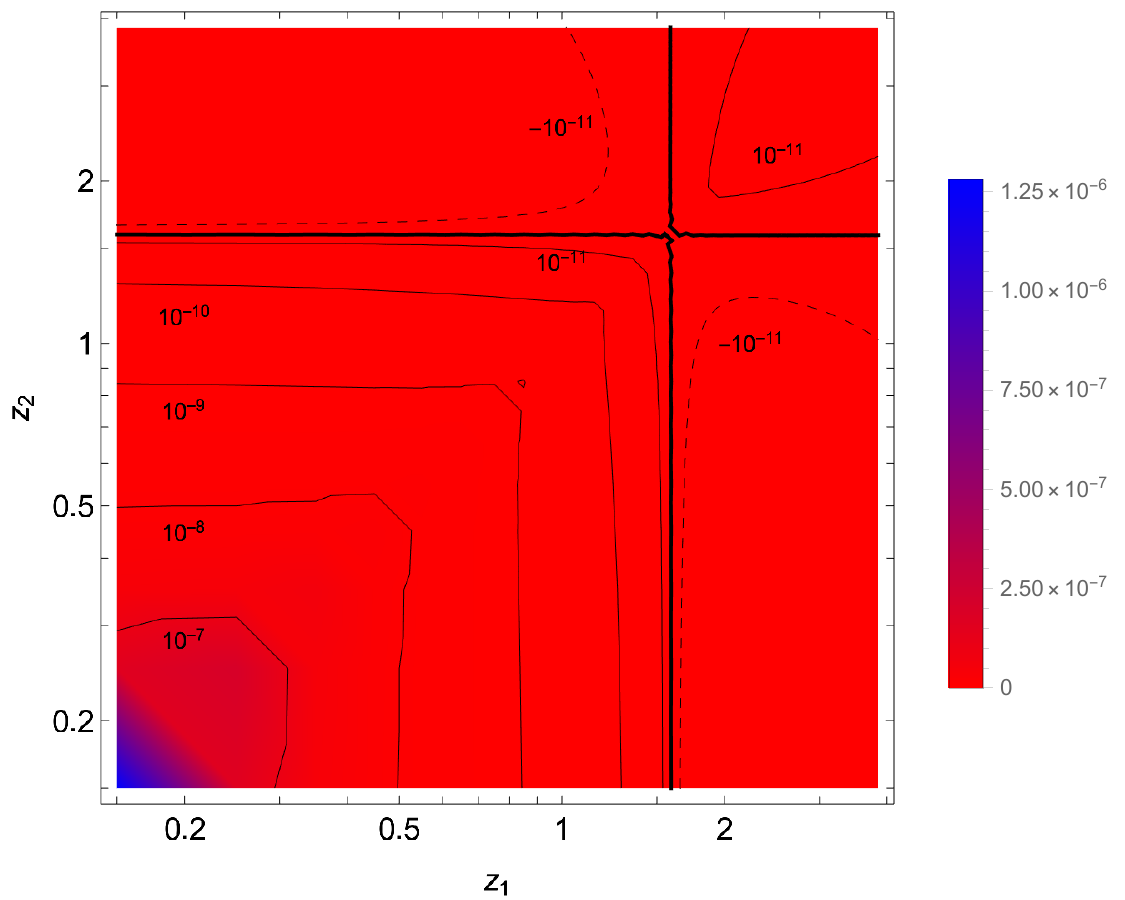}
\caption{Monopole of the doppler 2-point correlation function. Thick black lines indicate where function is 0. Dashed lines stand for negative values whereas continues lines refer to positive values.}
\label{fig:dopp_mon}
\end{figure}

Another interesting feature emerging from Eqs. \eqref{eq:pure_kernels} and \eqref{eq:mixed_kernels} regards the angular average of $\xi_{E,E'}$. It turns out that
\begin{equation}
\int_{-1}^{1}d\nu\,\xi_{E,L}(z_i,z_j,\nu)=0\,
\end{equation}
where $E$ may label all the relativistic effects such as lensing itself, peculiar velocity, time delay and (integrated) Sachs-Wolfe effect. This means that, in the limit of full sky coverage surveys and large number of sources\footnote{See Yoo (2020)\cite{Yoo:2019skw} for a detailed discussion of this ideal case.}, lensing does not affects at all the estimation $\sigma^2_{H_0}$. Hence, it follows that in this ideal case the leading contribution to the cosmic variance is entirely addressed to the monopole of $\xi_{PV,PV}$. This can be easily evaluated from the angular integration of the $\mathcal{W}_{PVi,PVj}$ in Eq. \eqref{eq:pure_kernels} and results are shown in Fig. \ref{fig:dopp_mon}.
In this ideal case, $\sigma_{H_0}$ is entirely given by the doppler 2-point correlation function.

However, realistic surveys deal with finite sky-coverages which can be also very narrow. In this case, monopole of the 2-point correlation functions of $\widetilde{d_L}(z,{\bf n})$ is no longer enough and higher multipoles must be taken into account for the evaluation of $\sigma^2_{H_0}$. We redirect the reader interested in the detailed multipoles analysis of $\xi_{E,E'}$ to Fanizza et al. (2021)\cite{Fanizza:2021tuh}. In the next section, we will apply the numerical results shown in Fig. \ref{fig:amplitudes} to the estimator for $\sigma^2_{H_0}$ in Eq. \eqref{DH_(W)}.

\section{Cosmic variance for Next Generation Surveys}
\label{sec:estimation}
\label{sec:next_gen_sur}
\begin{figure}
\centering
\includegraphics[scale=0.75]{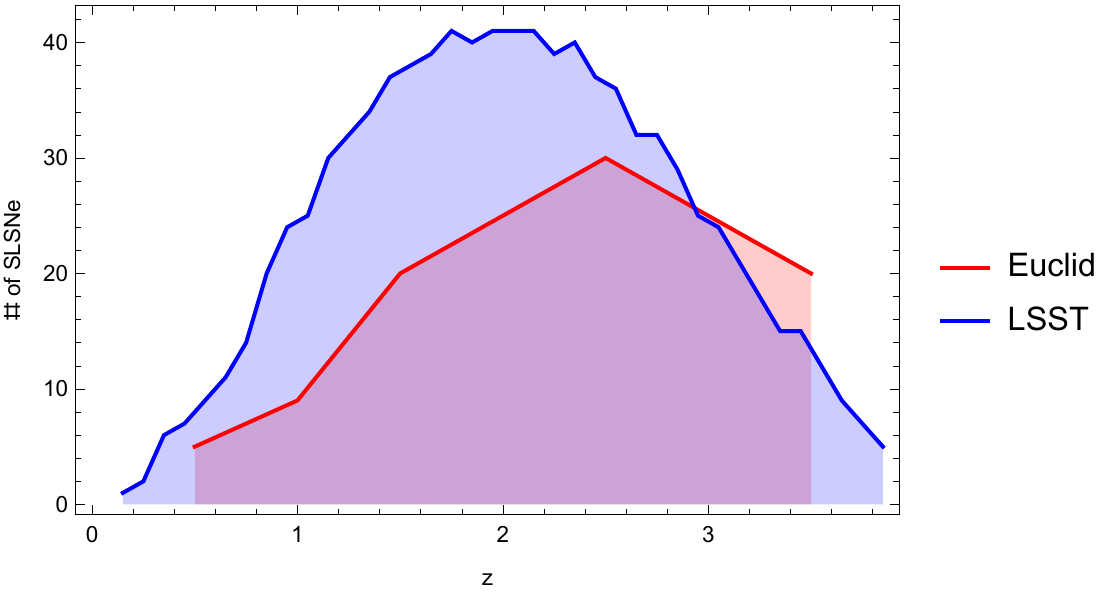}
\caption{Simulated SLSNe distributions for Euclid Deep Survey and LSST. From Inserra et al. (2021)\cite{Inserra:2020uki}.}
\label{fig:hist}
\end{figure}

In this section, we apply the analytical and numerical results previously obtained to the case of Superluminous Supernovae (SLSNe). Following the specifics of forthcoming surveys like EDS \cite{Laureijs:2011gra} and LSST \cite{Abell:2009aa}, we consider the expected detection rate of SLSNe claimed in Inserra et al. (2021)\cite{Inserra:2020uki} and reported in our Fig. \ref{fig:hist}.
With these histograms in mind, we have generated two random surveys for the distribution of SLSNe with the following properties
\begin{itemize}
\item EDS: 135 sources in 7 redshift bins with $0.5\le z \le 3.5$ and redshift bin width $\Delta z=0.5$. For this survey, the angular distribution covers two line-of-sights at North and South Poles, with angular opening of 20 deg$^2$ per line-of-sight,
\item LSST: 929 sources in 38 redshift bins with $0.15\le z \le 3.85$ and redshift bin width $\Delta z=0.1$. For this survey, the angular distribution spans a broad solid angle of 9000 deg$^2$.
\end{itemize}
Results are summarized in Table \ref{tab:results}.
\begin{table}
\tbl{Forecasts for the variance of $H_0$ in EDS and LSST. In the first line, lensing 2-point correlation function is considered. In the second line, the contributions from 2-point correlation function of peculiar velocities are shown. Values are reported for linear power spectrum truncated at $k_{UV}=0.1 h$ Mpc$^{-1}$ and Non-Linear HaloFit model truncated at $k_{UV}=10 h$ Mpc$^{-1}$.}
{\begin{tabular}{ccccc}
\toprule
$\sigma^2_{H_0}/H^2_0$&EDS (Linear)&LSST (Linear)&EDS (Non-Linear)&LSST (Non-Linear)\\
\colrule
Lensing&$5.1\times 10^{-6}$&$7.6\times 10^{-8}$&$1.1\times 10^{-5}$&$7.8\times 10^{-7}$\\
Doppler&$2.1\times 10^{-9}$&$2.9\times 10^{-10}$&-&-\\
\botrule
\end{tabular}}
\label{tab:results}
\end{table}
Here we see that the dispersion associated to the measure of $H_0$, namely $\sigma_{H_0}\equiv \sqrt{\sigma^2_{H_0}}$ is of $\sim 0.2 \%$ for EDS but its value drops of almost 1 order of magnitude for LSST, where it contributes with a dispersion of $\sim 0.03\%$. This difference is mainly addressed to the fact that, for larger sky coverage and large number of sources, the total effect due to lensing must tend to 0. The specific of LSST are indeed in line with regime: first of all, the number of sources adopted in our forecast of LSST is almost 1 order of magnitude higher than the one of EDS. Secondly, also the sky coverage is larger.

Another fact that we underline is that doppler effect is always subdominant. Hence we have that the total cosmic variance due to lensing and doppler is
\begin{equation}
\sigma_{H_0}
\approx\sigma_{H_0 L}\left[ 1+\frac{1}{2}\frac{\sigma^2_{H_0 PV}}{\sigma^2_{H_0 L}}+\mathcal{O}\left(\left(\frac{\sigma^2_{H_0 PV}}{\sigma^2_{H_0 L}}\right)^2\right) \right]\,.
\end{equation}
Again from Table \ref{tab:results}, we then get that the doppler effect corrects the total cosmic variance associated to $H_0$ by $0.2\%$ for LSST and by $0.02\%$ for EDS.

This analytical estimation can be naturally extended to $k_{UV}=10\, h$ Mpc$^{-1}$ where the HaloFit model\cite{Smith_2003,Takahashi_2012} is taken into account to model the non-linear scales. Results for the lensing are reported again in Table \ref{tab:results} and show that $\sigma^2_{H_0\,NL}=1.1\times 10^{-5}\,H^2_0$ for EDS and $\sigma^2_{H_0\,NL}=7.8\times 10^{-7}\,H^2_0$ for LSST.

It follows then, for EDS, even if non-linear scales are expected to enhance the lensing correction by almost one order of magnitude, the intrinsic error associated to the measurement of $H_0$ is almost insensitive to the non-linear scales, since it becomes $\sigma_{H_0\,NL}/H_0=0.003$. On the contrary, non-linear scales increase by roughly a factor 3 the dispersion of $H_0$ within the specific of LSST, raising $\sigma_{H_0}$ to the value $\sigma_{H_0\,NL}/H_0=0.0009$.

Our analysis, provides also a test for the claim done in Ben-Dayan et al. (2014)\cite{Ben-Dayan:2014swa} about small redshift surveys: it has been stated there that the analysis is insensitive to smaller scales fluctuations due to the incoherence of such contributions. According to our investigation, we conclude that this is a reasonable expectation only for EDS. We address this feature to the fact that EDS covers smaller regions in the sky with higher angular density of sources.

%%%%%%%%%%%%%%%%%%%%%%%%%%%%%%%%%%%%%%%

\section{Summary and conclusions}
\label{sec:discussion}
\label{sec:conclusions}
In this work, we have studied the impact of cosmological inhomogeneities on the estimation of $H_0$ from the high redshift Hubble diagram. Our analysis considers the possibility, discussed in Inserra et al. (2021)\cite{Inserra:2020uki}, that EDS \cite{Laureijs:2011gra} and LSST \cite{Abell:2009aa} will have access to a statistically relevant number of Superluminous Supernovae in the next decade. In this regards, less conservative studies about the Hubble diagram at high redshifts ($z\le 1.5$) have been also investigated by exploiting exact inhomogeneous models in general relativity \cite{Cosmai:2013iga,Romano:2016utn,Cosmai:2018nvx,Vallejo-Pena:2019agp} or by considering strongly inhomogeneous dynamical dark energy models \cite{Cai:2021wgv}. These attempts look interesting especially in light of recent analysis done in Krishnan et al. (2020)\cite{Krishnan:2020obg} and Krishnan et al. (2021)\cite{Krishnan:2020vaf} about H0LiCOW\cite{Wong:2019kwg} and TDCOSMO\cite{Millon:2019slk} data and in Dainotti et al. (2021)\cite{Dainotti:2021pqg} for the SNe Ia Pantheon sample, suggesting that $H_0$ could be a decreasing function of redshift already at late time.

However, our approach is more conservative, since it is based on linear perturbations within the Cosmic Concordance model. In this framework, the 2-point correlation function of luminosity distance-redshift relation has been analytically derived and investigated numerically. It turns out that lensing is the leading effect at the considered redshift, as one may expect.

Our first analytical estimations of the cosmic variance for limited-sky-coverage surveys indicate then that forthcoming high-redshift surveys are well-suited to provide a precise determination of cosmological parameters, such as $H_0$.
Moreover, these forecasted errors for LSST and EDS are stable enough to be quite insensitive to the role of non-linearities in the matter power spectrum.
Despite our analysis has been performed entirely within the $\Lambda$CDM model, from the geometrical structure of the light-cone 2 general features emerge:
\begin{itemize}
\item Lensing 2-point correlation function has vanishing monopole, hence in the limit of large sky coverage and huge number of sources, the cosmic variance must be dominated by Doppler effect also on high redshift surveys.
\item As a consequence of line-of-sight integration and statistic isotropy, non-linear scales play an important role only when two sources are almost aligned.
\end{itemize}

However, realistic surveys deal with limited sky coverage and this makes lensing contribution no longer vanishing. In fact, according to the specific of EDS and LSST and to what has been claimed in Inserra et al. (2021)\cite{Inserra:2020uki}, we forecast that the intrinsic error from cosmic variance associated to $H_0$ is of $\sim0.03\,\%$ for LSST and $0.3\,\%$ for EDS for the linear power spectrum. Non-linear scales contribute marginally to this estimation within the specific of EDS. For what regard the specific of LSST, the situation is a way worse. Indeed, in this case, we get that non-linear scales enhance our forecast by almost a factor 3. This is a direct consequence of the fact that lensing 2-point correlation function strongly depends on small scales fluctuations for the diagonal entries of the covariance matrix.

Our analysis extends the one performed in Ben-Dayan et al. (2014)\cite{Ben-Dayan:2014swa}, where only close Supernovae (up to $z=0.1$) have been considered. Indeed, we took into account also lensing corrections on top of the peculiar motion of the sources. In fact, the former is the leading relativistic effect expected at those redshifts\cite{BenDayan:2013gc}. Interestingly, our analysis points out that surveys have an intrinsic error for $H_0$ which tends to decrease when higher redshift sources are considered, whereas low redshift surveys discussed in Ben-Dayan et al. (2014)\cite{Ben-Dayan:2014swa} admits a quite high dispersion for $H_0$ of $\sim 1\,\%$.

On one hand, our results are not able to alleviate the tension between local and distant measurements of the Hubble constant. However, they indicate that the analysis of fainter sources does not increase the theoretical uncertainty on $H_0$. The price to pay stands in the fact that the Hubble diagram at higher redshift is no longer model independent.

\bibliographystyle{ws-procs961x669}
\bibliography{biblio}

\end{document}